# High-pressure polymeric nitrogen allotrope with the black phosphorus structure

Dominique Laniel[1*], Bjoern Winkler[2], Timofey Fedotenko[1], Anna Pakhomova[3], Stella Chariton[4], Victor Milman[5], Vitali Prakapenka[4], Leonid Dubrovinsky[6], Natalia Dubrovinskaia[1,7].

**Affiliations:**

[1]Material Physics and Technology at Extreme Conditions, Laboratory of Crystallography, University of Bayreuth, 95440 Bayreuth, Germany.

[2]Institut für Geowissenschaften, Abteilung Kristallographie, Johann Wolfgang Goethe-Universität Frankfurt, Altenhöferallee 1, D-60438, Frankfurt am Main, Germany.

[3]Photon Science, Deutsches Elektronen-Synchrotron, Notkestrasse 85, 22607 Hamburg, Germany.

[4]Center for Advanced Radiation Sources, University of Chicago, Chicago, Illinois 60637, United States

[5]Dassault Systèmes BIOVIA, CB4 0WN Cambridge, United Kingdom

[6]Bayerisches Geoinstitut, University of Bayreuth, 95440 Bayreuth, Germany.

[7]Department of Physics, Chemistry and Biology (IFM), Linköping University, SE-581 83, Linköping, Sweden

*Correspondence to: dominique.laniel@uni-bayreuth.de.

**Abstract**

Studies of polynitrogen phases are of great interest for fundamental science and for the design of novel high energy density materials. Laser heating of pure nitrogen at 140 GPa in a diamond anvil cell led to the synthesis of a polymeric nitrogen allotrope with the black phosphorus structure, bp-N. The structure was identified *in situ* using synchrotron single-crystal X-ray diffraction and further studied by Raman spectroscopy and density functional theory calculations. The discovery of bp-N brings nitrogen in line with heavier pnictogen elements, resolves incongruities regarding polymeric nitrogen phases and provides insights into polynitrogen arrangements at extreme densities.

**Introduction**

The behavior of simple diatomic molecular solids under high pressure is one of the fundamental problems of condensed matter physics, and their experimental investigation provides benchmarks for modern theories of the solid state. Despite its apparent simplicity, molecular nitrogen displays a complex behavior at high densities as it possesses the strongest homoatomic covalent bond, the shortest bond length and a significant electric quadrupole moment. The electronic quadrupole-quadrupole interaction between $N_2$ molecules increases sharply as intermolecular distances decrease and thus largely dictates the low to moderate pressures (~ 5 to 50 GPa) behavior of nitrogen [1]. At higher pressures (> 50 GPa), as the electrons' kinetic energy surpasses the electrostatic potential energy, part of the molecules' triple bond electronic density is shifted to intermolecular regions [2,3]. Such an electron density delocalization leads to a progressive weakening of the N≡N intramolecular triple bond [2,4]. Due to these intricate processes, up to 110 GPa nitrogen exhibits an impressive polymorphism, with 11 experimentally observed crystalline phases [5]. Above 110 GPa, intramolecular bonds in $N_2$ are weakened to the extent that they are ruptured at temperatures of ~ 2500 K, and a polymeric network of single-bonded nitrogen atoms is formed [6]. Hitherto, nitrogen has been experimentally investigated up to 244 GPa and 3300 K, and three crystalline polymeric phases have been discovered: *cubic-gauche* polymeric nitrogen (cg-N), layered polymeric nitrogen (LP-N), and hexagonal layered polymeric nitrogen (HLP-N), synthesized at high temperatures and at 110-180 GPa, 125-180 GPa, and 244 GPa, respectively [6–8]. While the crystal structures of cg-N and HLP-N are well established, some doubts have been raised regarding the LP-N structure [8,9].

Polymeric nitrogen phases are of great interest as prototypes for the design of novel high energy density materials [10–12], which led to intensive theoretical studies and numerous crystal structure predictions [9,13–27]. Among these, the layered polymeric structure of black phosphorus—the thermodynamically stable form of phosphorus at ambient conditions [28]—has been suggested in earlier theoretical works as a potential candidate for polymeric nitrogen [13,15–17]. This structural prediction was partially inspired by the commonly accepted paradigm that elements of the same periodic table column adopt the structure of the elements below them, but at higher pressure [29]. This empirical model of the structural behavior of elements under pressure generally holds for the nitrogen family group elements (pnictogens) [30], with an exception for nitrogen itself, as up to now nitrogen has not been shown to adopt the black phosphorus structure.

In this manuscript, we present high pressure and high temperature experiments on nitrogen up to 140 GPa and ~4000 K. We performed synchrotron single-crystal X-ray diffraction on polycrystalline samples and unambiguously identified a polymeric nitrogen allotrope with the black phosphorus structure (bp-N). Complementary Raman spectroscopy measurements and density functional theory (DFT) calculations enabled us to unambiguously establish that the previously reported LP-N structure [7] is incorrect, and that this solid in fact adopts the bp-N structure. The discovery of the bp-N structure resolves previous anomalies in the density evolution of the polymeric nitrogen phases under pressure and emphasizes the favorable nature of high density polynitrogen arrangement to form 2D layered systems composed of distorted and fused $N_6$ rings. The crystal structure of bp-N is also discussed in analogy with black phosphorus and black arsenic—exposing the N-N bonds' disparity in electronic density and lengths, thus leading to an enhancement of the structural anisotropy characteristic of the black phosphorus structure

**Experimental Method, Results and Discussion**

A BX90-type diamond anvil cell [31] equipped with 80 μm diameter diamond anvil culets was prepared. A 200 μm thick rhenium foil was indented down to 12 μm and a sample cavity of 40 μm in diameter was laser-drilled at the center of the indentation. Two agglomerates of submicron-sized gold particles, each of approximately 2 μm in size, were loaded into the sample chamber to serve as both YAG laser absorbers and pressure gauges [32]. The cell was then loaded with pure nitrogen gas at ~1200 bars. The sample was compressed to 124 GPa and laser-heated, reaching a maximum temperature of 2600 K as determined by thermoemission measurements [33]. Weak diffraction spots appeared after laser heating but could not be indexed using any of the known phases of molecular or polymeric nitrogen, gold, or rhenium. The laser-heated portion of the sample transformed from mostly dark, as expected for nitrogen at this pressure [34–36], to transparent, as seen in Figure S1. To promote further conversion into this new solid and to obtain higher quality diffraction data, the pressure was increased to 140 GPa and nitrogen was reheated, this time to temperatures of ~4000 K. Significantly more diffraction spots with higher intensities were then detected. A single-crystal X-ray diffraction data collection was performed on the polycrystalline sample and the resulting dataset was analyzed following an established procedure [11,37,38], allowing for an unambiguous structural characterization.

The high pressure nitrogen allotrope synthesized here is isostructural to black phosphorus [39]. The bp-N structure, shown Figure 1, is orthorhombic (space group *Cmce*) with a single Wyckoff position (8*f*) occupied by nitrogen atoms (see Table 1). The structure is layered, and each stratum is constituted of three-fold coordinated single bonded nitrogen atoms. A monolayer may be regarded as a system of interconnected zigzag (ZZ) chains that alternate in two different planes. These ZZ chains are linked together through an armchair (AC) arrangement running along [101] direction. At 140 GPa, the N-N bond length (angle) in the ZZ and AC layouts are 1.338(6) Å (105.8(7)°) and alternate between 1.338(6) and 1.435(7) Å (105.19(10)°), respectively. Alternatively, the nitrogen atom network can also be understood as a puckered honeycomb arrangement, *i.e.* distorted and fused $N_6$ hexagons with four shorter and two longer bonds, as well as with four ∠N-N-N angles of 105.16(10)° and two of 105.6(7)°. The shortest interlayer N-N distance is of 2.329(6) Å, effectively prohibiting the formation of interlayer covalent. DFT calculations (see Supplementary Materials for details) found this structure dynamically stable and the relaxed structural parameters, also shown in Table *1*, closely reproduce the experimental values.

A Le Bail analysis of the powder X-ray diffraction patterns yielded the unit cell and space group of the new bp-N phase. As seen in Figure S2, all diffraction lines could be attributed to one of three phases: bp-N, Re or Au. No signature of cg-N was detected, neither from X-ray diffraction nor from Raman spectroscopy. This is in contrast to previous experiments in which cg-N was always detected after heating nitrogen at pressures between 125 and 180 GPa [7]. From another pure nitrogen sample compressed to 122 GPa and laser-heated, the structure of cg-N was successfully solved by single-crystal X-ray diffraction. The complete crystallographic details can be found in Table S1, and validate the literature data [6,40]. Notably, the distance between single bonded nitrogen atoms in cg-N at 122 GPa is of 1.341(3) Å; in good agreement with the shortest bond length in bp-N (1.338(6) Å at 140 GPa).

**Table 1: Experimentally determined crystallographic data for the bp-N phase at 140 GPa in comparison with the corresponding DFT-relaxed structure. The full crystallographic dataset was deposited to the CCDC under the deposition number 1986002.**

| Crystal phase | bp-N (experiment) | bp-N (theory) |
|---|---|---|
| Pressure (GPa) | 140 | 140 |
| Space group | *Cmce* | *Cmce* |
| $a$ (Å) | 2.1343(7) | 2.1459 |
| $b$ (Å) | 6.534(2) | 6.5655 |
| $c$ (Å) | 2.860(4) | 2.8297 |
| $V$ (Å$^3$) | 39.88(6) | 39.87 |
| Fractional atomic coordinates ($x$; $y$; $z$) | N: (0; 0.4011(4); 0.1090(17)) | N: (0; 0.3899; 0.1001) |
| Number of measured/independent reflections (I ≥ 3$\sigma$) | 101 / 31 (22) | |
| $R_{int}$ | 0.0879 | |
| Final R indexes (I ≥ 3$\sigma$) | $R_1$ = 0.0592; $wR_1$ = 0.0635 | |
| Final R indexes (all data) | $R_1$ = 0.0760; $wR_1$ = 0.0697 | |
| Number of refined parameters | 4 | |

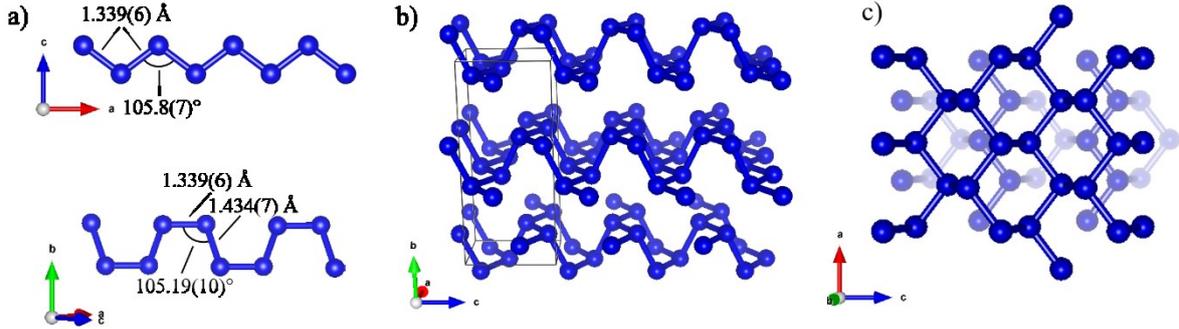

**Figure 1:** a) The zigzag (ZZ) and armchair (AC) arrangements forming the bp-N layers. b) The crystal structure of bp-N, where the ZZ and AC chains are visible. c) Two superimposed layers of the bp-N structure, highlighting the puckered honeycomb arrangement of fused $N_6$ hexagons.

Raman spectroscopy of bp-N showed two very intense vibrational modes at frequencies of 1001 and 1308 cm$^{-1}$ at 140 GPa, suspiciously similar to those previously thought to belong to LP-N [7,8]. This incongruity motivated the DFT calculation of the Raman modes associated with the bp-N and LP-N structures. It is well established that such lattice dynamical calculations reproduce frequencies and intensities to within a few percent [41–45]. The comparison between the calculated spectrum of bp-N and LP-N as well as the experimental spectrum is shown in Figure 2. Strikingly, the modes calculated for the LP-N structure fail to correctly reproduce the modes measured experimentally, while those of the calculated bp-N structure perfectly matches them. Moreover, while a factor group analysis shows that there are 45 Raman active modes ($\Gamma = 11A_1 + 12A_2 + 11B_1 + 11B_2$) for LP-N [7], there are only six ($\Gamma = 2A_g + B_{1g} + B_{2g} + 2B_{3g}$) Raman active modes in bp-N. Five modes have been experimentally identified [7,8]. The sixth calculated mode ($B_{2g}$) of bp-N, which is not visible in the recorded spectra, is expected to be of low intensity and at a frequency of 1454 cm$^{-1}$; hence overlapping with diamond's first order Raman mode which extends from 1330 cm$^{-1}$ up to about 1610 cm$^{-1}$ at 140 GPa. For the assignment of irreducible representations to the five measured Raman peaks (Figure 2), DFT calculations were employed. In summary, these results unambiguously demonstrate that the polynitrogen polymorph observed in previous experiments has the bp-N structure described in this work.

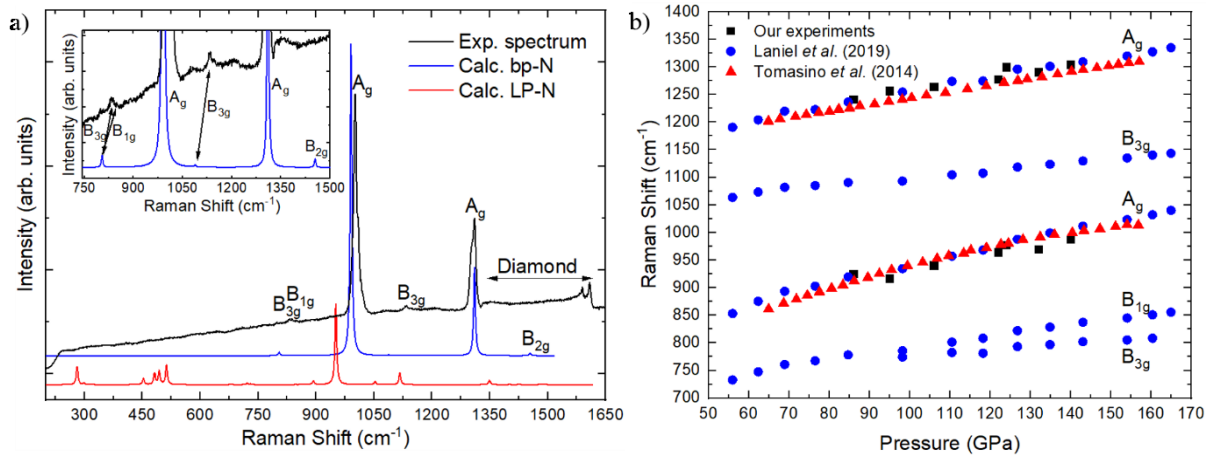

**Figure 2:** a) Comparison between the experimentally measured modes and those calculated for LP-N and bp-N at 140 GPa. (Inset) An enlargement of the 750 to 1500 cm$^{-1}$ frequency range, allowing to better see the very weak vibrational modes of the bp-N structure: $B_{1g}$, the two $B_{3g}$ and $B_{2g}$, at the calculated frequencies of 804, 804, 1088 and 1454 cm$^{-1}$, respectively. For clarity, arrows indicate the correspondence between the calculated and experimental low intensity modes. b) Pressure dependence of the Raman modes obtained here compared to those previously attributed to the LP-N structure, now established to belong to bp-N [7,8].

The bp-N sample was decompressed down to 86 GPa—the pressure at which it escaped from the sample cavity—while being characterized by Raman spectroscopy and X-ray diffraction. The Raman modes measured during the pressure decrease also closely match those previously attributed to the LP-N phase, as seen in Figure 2 b) [7,8]. Due to the deteriorating quality of the single crystals upon decompression, only the pressure-dependence of the unit cell parameters could be obtained from powder X-ray diffraction data. A comparison of the volume per nitrogen atom in the cg-N, bp-N and HLP-N structures provides a coherent picture in which a volume collapse of about 1% and 2% is detected for the cg-N → bp-N and the bp-N → HLP-N phase transitions, respectively (Figure 3). For the erroneous LP-N structure, instead, an enormous volume collapse of ~7% compared to cg-N was reported [7], suggesting in turn a suspicious volume increase between LP-N and the higher pressure phase HLP-N [8]. Pressure-volume data obtained from DFT calculations reproduce the experimental data and corroborate the volume-decreasing trend from cg-N → bp-N → HLP-N.

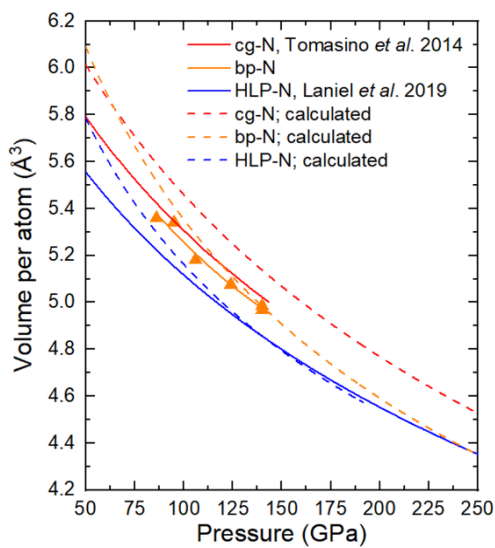

**Figure 3: Experimental and DFT calculated volume per nitrogen atom in function of pressure for cg-N, bp-N and HLP-N [7,8]. The orange triangles represent experimental data obtained here on bp-N, while the full and dashed lines are fits to the experimental and calculated data, respectively. Notably, the volume per nitrogen atom decreases when transforming from cg-N to bp-N, as well as from bp-N to HLP-N.**

With the structures of all known polymeric nitrogen allotropes firmly established, accurate comparisons between their respective nitrogen networks can be achieved. First, the pressure-induced shift between 3D → 2D arrangements from cg-N to bp-N and HLP-N is noticed, and the favorable energetics of a layered structure as opposed to a 3D framework at higher densities is now unambiguously established. This drop in dimensionality, while surprising, was also observed in the polymeric form of CO and is expected from polymeric $CO_2$ [46,47]. The two stratified compounds bp-N and HLP-N are alike, both being composed of distorted and fused $N_6$ rings—hypothetically underlining the advantageous nature of such arrangements at high pressures to accommodate the increase of electronic density. The interlayer gap distance decreases from 2.329 Å in BP-N at 140 GPa down to only 2.02 Å in HLP-N at 235 GPa; in the latter case approaching the distance of intermolecular voids in molecular $N_2$ at 100 GPa [2,8]. Calculated electron localization function maps (see Figure S3) display no charge accumulation at mid-points between lines connecting N-N atoms in neighboring layers, analogously to HLP-N [8]. These interlayer gaps are accommodating the stereochemically active lone electron pairs from all nitrogen atoms.

Enthalpy calculations for bp-N, LP-N as well as cg-N between 130 and 180 GPa were performed and their result are shown in Figure S4. Interestingly, the bp-N and LP-N structures are within a few meV/atom of each other, with the former marginally more favorable below 150 GPa. However, in agreement with the latest enthalpy calculations [20], cg-N appears to be the most thermodynamically stable phase up to at least 180 GPa by a small margin of 10-80 meV/atom. Fully deciphering the reasons for the formation of bp-N instead of the enthalpically preferred cg-N solid is beyond the scope of the present work. However, the current model calculations neglect temperature effects, and we did not investigate the effect of activation barriers. It is noteworthy to point out that similar concerns were raised for the HLP-N solid, which is also calculated—both in the athermal limit and at 3000 K—not to be the lowest enthalpy phase at any pressures, despite being synthesized [8]. This stresses the importance of techniques enabling structure solutions fully independent of theoretical calculations, such as the recently developed methodology of single crystal X-ray diffraction from polycrystalline samples [38], used in the present work.

The exotic thermal, electronic, and optical properties of black phosphorus are thought to originate due to the distinct ZZ and AC P-P chains' geometry in this anisotropic orthorhombic crystal structure [48–50]. In spite of the radical difference of physical properties along the ZZ and AC directions, the P-P bond lengths in these chains are practically identical, ranging between 2.224 and 2.262 Å (for a bond length difference of 1.7%) [39]. Arsenic can also form the same structure type as black phosphorus, referred to as black arsenic, although this solid is not a thermodynamic ground state at ambient conditions or under pressure [51]. Black arsenic also exhibits unique physical properties and, analogously to black phosphorus, it is displaying very similar As-As bond lengths, namely of 2.435 Å and of 2.428 Å (bond length difference of 0.3%) for the ZZ and AC contacts [51]. In the case of bp-N however, very dissimilar values of 1.338(6) and 1.435(7) Å (bond length difference of 7.2%) were measured as N-N interatomic distances in the ZZ and AC chains, respectively. According to the computed electron difference maps (see Figure S3), this disparity in bond lengths translates into a significantly larger bond population observed between neighbors along the ZZ chains (1.26 $e^-$/Å$^3$), about seven times the bond population calculated along the AC chains (0.18 $e^-$/Å$^3$). The consequences of the N-N bond difference are expected to exacerbate the compound's structural anisotropy.

In agreement with the experimentally observed transparency of bp-N, electronic band structure calculations, shown in Figure S5, suggest that it is a wide bandgap semiconductor (2.2 eV at 150 GPa), in contrast to the extremely narrow bandgap (0.3 eV) of black phosphorus and black arsenic [51,52]. This is not unexpected as within a given column of the periodic table—and particularly within groups 13 to 17— the lower $Z$ elements typically have a higher band gap than their high $Z$ counterparts [53]. A similar trend is also observed in many compounds, for example in zinc-blende type semiconductors [54,55].

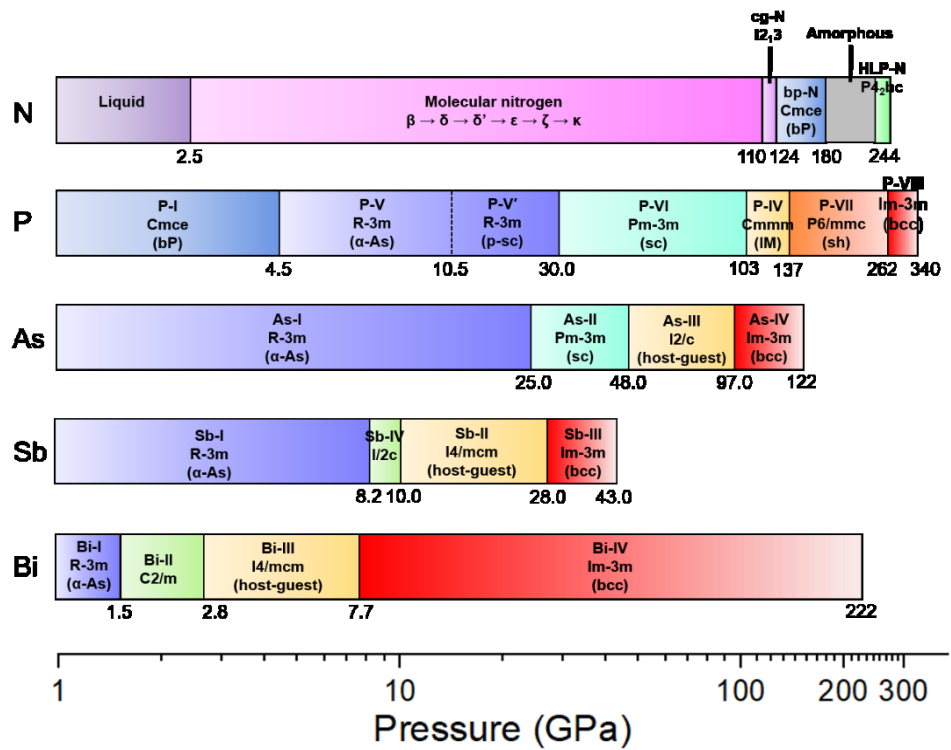

Figure 4: Schematic diagram illustrating the stable phases of the pnictogen elements with pressure. When possible, each phase is identified with its commonly employed name, space group and structure type. The terms p-sc, sc, IM, bcc and sh refer to pseudo-simple cubic, simple cubic, incommensurate, body centered cubic and simple hexagonal, respectively. Phases indicated with the same color possess the same structure type. The figure was modified after references [56] and [57].

The synthesis of bp-N also brings nitrogen in line with higher-Z pnictogen elements, and reaffirms the trend that elements at high pressure adopt the same structures as the same-group elements below them in the periodic table at lower pressures [29,58]. Indeed, until now, none of the numerous known nitrogen structures matched those adopted by the pnictogen-group elements, which were at least sharing the same rhombohedral *R-3m* layered structure (α-As) [56]. The similarities between N and P, As, Sb and Bi could further endure if nitrogen is found to also form the α-As type structure at higher pressures. In particular, further investigations of polymeric nitrogen between 180 and 244 GPa, pressure at which no crystalline phases were thus far detected [8], and at higher temperatures could favor the synthesis of other polymeric phases known to pnictogen elements.

**Conclusions**

Here, pure molecular nitrogen was investigated up to 140 GPa and laser-heated to produce a previously unknown polymeric nitrogen allotrope. Through single crystal X-ray diffraction measurements of the resulting polycrystalline sample, the synthesized solid was unambiguously shown to be isostructural to black phosphorus (space group *Cmce*). The Raman spectrum of bp-N was found to perfectly reproduce the experimental vibrational spectrum previously assigned to the LP-N phase. This finding was unambiguously confirmed by DFT calculations, which showed that the Raman spectra of bp-N and LP-N can easily be distinguished from one another and where only the spectra computed for bp-N agrees with the experimental observations. These results thus unambiguously demonstrate that the LP-N polymorph observed in previous experiments has the bp-N structure described in this work.

Having now firmly established the crystal structure of cg-N, bp-N and HLP-N, $N_6$-bearing layered structures appear as the thermodynamically favored arrangement of polynitrogen at high densities. The DFT calculations point towards an enhanced anisotropy along the ZZ and AC directions of bp-N due to significant differences in N-N bond lengths and electron densities between neighboring atoms. The calculations also identified bp-N as the sole semiconducting phase of nitrogen with a wide bandgap of about 2.2 eV. The structural identification of this nitrogen polymeric phase also reconciles the pnictogen elements, now with nitrogen and phosphorus having the same 2D layered orthorhombic crystal structure. This study also enables a better understanding of the behavior of homoatomic polymeric solids under pressure and resolves previous incongruities in the density evolution of polymeric nitrogen phases. Further investigations into the electronic, optical and thermal properties of bp-N—the structural analog of black phosphorus—are expected.

**Acknowledgments**

The authors acknowledge the Deutsches Elektronen-Synchrotron (DESY, PETRA III) and the Advanced Photon Source (APS) for provision of beamtime at the P02.2 and GSECARS beamlines, respectively. Alexander Kurnosov is greatly thanked for helping with sample loadings employing the high pressure gas loader. Jinke Bao is also acknowledged for helpful discussions. D.L. thanks the Alexander von Humboldt Foundation for financial support. N.D. and L.D. thank the Federal Ministry of Education and Research, Germany (BMBF, grants no. 5K16WC1 and no. 05K19WC1) and the Deutsche Forschungsgemeinschaft (DFG projects DU 954–11/1, DU 393–9/2, and DU 393–13/1) for financial support. B.W. gratefully acknowledges funding by the DFG in the framework of the research unit DFG FOR2125 and within projects WI1232. N.D. thanks the Swedish Government Strategic Research Area in Materials Science on Functional Materials at Linköping University (Faculty Grant SFO-Mat-LiU No. 2009 00971).